# Threefold error in the reported zero-field cooled magnetic moment of single crystal La$_2$SmNi$_2$O$_7$

Aleksandr V. Korolev[*] and Evgeny F. Talantsev[**]

M.N. Miheev Institute of Metal Physics, Ural Branch, Russian Academy of Sciences,
18, S. Kovalevskaya St., Ekaterinburg, 620108, Russia

* korolyov@imp.uran.ru
** evgeny.talantsev@imp.uran.ru

**Abstract**

For a relatively long time, the observation of the DC diamagnetic state in highly compressed nickelate superconductors[1], [2] has been a challenging experimental problem. And recently Li et al.[3] reported on the measurements of the DC diamagnetism in zero-field-cooled (ZFC) and field-cooled (FC) pressurized single crystal La$_2$SmNi$_2$O$_7$. From the analysis of experimental data, Li et al.[3] reported that the superconducting phase fraction in their La$_2$SmNi$_2$O$_7$ sample measured in the ZFC mode is 62.1%, and the superconducting phase fraction in the FC mode is 14.4%. It should be clarified that we regard the measurements of the DC diamagnetic state[3] in La$_2$SmNi$_2$O$_7$ (and more recently[4] in Pr$_4$Ni$_3$O$_{10}$) as outstanding experimental results confirming bulk superconductivity in pressurized Ruddlesden-Popper nickelates. However, we should note that Li et al.[3] made a threefold error in their calculations of the superconducting phase fraction in La$_2$SmNi$_2$O$_7$. We believe that correcting this and other errors in Ref. [3] will benefit the physics community.

Our first note is that the FC data cannot be used for any superconducting phase fraction calculations, because of the paramagnetic Meissner effect[5], [6], [7], [8] (also known as the Wohlleben effect[5], [6], [7], [8]), which implies that the magnetic moment of a genuine superconducting sample in the FC mode can be as positive as negative. However, authors[3] calculated the superconducting phase fraction based on the FC data.

Our second note is that our analysis of ZFC data[3] showed that the calculation approach used by Li et al.[3] should yield a value three times smaller, 22.8%, than the reported 62.1% by the authors[3]. We presented these calculations below.

Third, we note that the approach used by Li et al.[3] to calculate the superconducting phase fraction from ZFC data is incorrect, since there are an infinite number of combinations of superconducting phase sizes smaller than the physical dimensions of the sample (and hence there is a fundamental uncertainty in the superconducting phase volume), resulting in the same magnetic moment as the measured moment in the ZFC mode. We presented the explanation of this issue below.

Let us first show that the values reported by Li et al.[3] and those we calculated using the approach used by Li et al.[3] differ by a factor of 3.

Li et al.[3] reported that the La$_2$SmNi$_2$O$_7$ single crystal (on which the ZFC and FC measurements were performed) had a disk-shaped shape with a diameter $d = 180$ μm and a thickness $h = 20$ μm. It should be noted that Li et al.[3] did not present the micrograph of this sample, while the micrographs for the other eighteen La$_{3-x}$Sm$_x$Ni$_2$O$_7$ samples, which were not studied by ZFC and FC magnetization, are given in Ref.[3].

For a superconductor in the Meissner state, the magnetic flux density inside the sample is zero by definition, $B = 0$. Thus, for a sample with demagnetization factor[9] $N$:

$$B = 0 = \mu_0 \times (M_V + H - N \times M_V), \tag{1}$$

$$M_V = -\frac{H}{1-N}, \tag{2}$$

where $\mu_0$ is the magnetic permeability of free-space (in units of $[NA^{-2}]$), $H$ is the applied magnetic field (in units of $\left[\frac{A}{m}\right]$), and $M_V$ is the sample volume magnetization (in units of $\left[\frac{A}{m}\right]$), which is calculated from the measured sample magnetic moment $m$ (in units of $[Am^2]$), and the volume of the sample $V$ (in units of $[m^3]$):

$$M_V = \frac{m}{V}. \tag{3}$$

Based on Eqs. 1-3, the magnetic moment of the sample $m$ (measured in a magnetometer) in the Meissner state is:

$$m = -V \times \frac{H}{1-N}. \tag{4}$$

Detailed description of this approach can be found in Ref.[10].

The demagnetization factor, $N$, for a disk shape sample can be calculated using the equation derived by Brandt[9]:

$$N = 1 - \frac{1}{1+q \times \frac{d}{h}}, \tag{5}$$

$$q = \frac{4}{3\pi} + \frac{2}{3\pi} \times tanh\left(1.27 \times \frac{h}{d} \times ln\left(1 + \frac{d}{h}\right)\right) \tag{6}$$

The substituting sample dimensions in Eqs. 5,6 yields $N = 0.81548$. Li et al.[3] used approximated equation, and their calculations yielded $N = 0.849$. Based on our $N$ value (Equations 5,6), the magnetic moment of the single crystal $La_2SmNi_2O_7$ in the Meissner state in the applied field $H = 795.77 \frac{A}{m}$ is:

$$m_{calc,Meissner}[Am^2] = -V[m^3] \times \frac{H\left[\frac{A}{m}\right]}{1-N} = -\left(\frac{\pi}{4} \times (1.8 \times 10^{-4})^2 \times (2 \times 10^{-5})\, m^3\right) \times$$

$$\frac{795.77 \frac{A}{m}}{1-0.81548} = -2.195 \times 10^{-9}\, Am^2. \tag{7}$$

Considering that the measured magnetic moment in ZFC mode reported by Li et al.[3] in their Fig. 3,a[3] (after the subtraction of the background) is:

$$m_{meas,ZFC}(P = 20.6\, GPa, T = 40\, K) = -5 \times 10^{-10}\, Am^2, \tag{8}$$

one can calculate the ratio (which was used by Li et al.[3] as the fraction of the superconducting phase in the sample):

$$\frac{m_{meas,ZFC}(P=20.6\, GPa, T=40\, K)}{m_{calc,Meissner}} = \frac{5 \times 10^{-10}\, Am^2}{2.195 \times 10^{-9}\, Am^2} = 0.2277 \cong 22.8\% \tag{9}$$

The obtained ratio is 22.8% (Eq. 9) and it is about three times lower than the value of 62.1% reported by Li et al.[3].

We should stress that we do not question the existence of the diamagnetic state or the diamagnetic measurements reported by Li et al.[3], and from our view, data reported by Li et al.[3] show that the superconductivity and the diamagnetism (as one of the primary physical manifestations of the superconducting state) are genuine physical states in highly compressed $La_{3-x}Sm_xNi_2O_{7-\delta}$ samples[3], as well as these phenomena are in other highly compressed nickelates[1], [2], [3], [11], [12], [13] and highly compressed hydride superconductors[14], [15], [16], [17], [18].

However, we do not agree that the ratio used in equation 9 can (even if correctly calculated) serve as a measure of the fraction of the superconducting phase in any superconductor. This is explained by the fact that data measured in ZFC mode can only confirm that the sample is in a pure Meissner state. This confirmation occurs when the experimentally measured value of $m_{meas,ZFC}$ equals the calculated value of $m_{calc,Meissner}$ (according to Equation 4):

$$m_{meas,ZFC} = m_{calc,Meissner}. \qquad (10)$$

If the experiment shows that:

$$m_{meas,ZFC} < m_{calc,Meissner}, \qquad (11)$$

then the fraction of the superconducting phase in the sample is unknown. This is due to two fundamental issues, which can be understood from Equation 4:

1. If the experiment shows that $m_{meas,ZFC} < m_{calc,Meissner}$, then the volume of the superconducting Meissner phase, its shape and distribution in the sample is unknown;
2. Based on point #1, the demagnetization factor $N$ cannot be calculated, since $N$ is calculated for precisely known geometric parameters/dimensions of the sample under the assumption of an ideal diamagnetic response of the sample.

This statement can be demonstrated by the calculations of the magnetic moment of the disk shape sample with diameter $d = 121$ μm and thickness $h = 1.5$ μm. The volume of this sample is 3.4% of the sample volume studied by Li et al.[3] (with diameter $d = 180$ μm and thickness $h = 20$ μm). However, this sample in the Meissner state and the applied field $H = 795.77 \frac{A}{m}$ has the same magnetic moment as the magnetic moment measured by Li et al.[3]:

$$m_{calc,Meissner}[Am^2] = -V[m^3] \times \frac{H\left[\frac{A}{m}\right]}{1-N} = -\left(\frac{\pi}{4} \times (1.21 \times 10^{-4})^2 \times (1.5 \times 10^{-6})\, m^3\right) \times \frac{795.77 \frac{A}{m}}{1-0.97254} = -5 \times 10^{-10}\, Am^2, \quad (12)$$

where $N = 0.97254$ was calculated by Equations 5,6.

As we stated above, there are an infinite number of disk geometries with $d \leq 180\ \mu m$ and $h \leq 20\ \mu m$ that have the same $m_{calc,Meissner}$ as the $m_{meas,ZFC}$ measured by Li et al.[3]. More details can be found in Ref.[19].

Based on the above, our primary disagreement with authors[3] is that the superconducting phase fraction in any superconducting sample can be determined (or even estimated) by the use of any ratios similar to Equation 9.

In conclusion, we congratulate the authors[3], [4] on their outstanding experimental achievement in measuring the DC diamagnetism in Ruddlesden-Popper nickelates $La_2SmNi_2O_{7-\delta}$[3] and $Pr_4Ni_3O_{10}$[4].

**Author contributions**

AVK and EFT jointly conceived the work; AVK analysed data and performed the calculations; EFT confirmed the calculations and wrote the manuscript, which was revised by AVK.

**Acknowledgements**

The work was carried out within the framework of the state assignment of the Ministry of Science and Higher Education of the Russian Federation for the IMP UB RAS.

**Competing interests**

The authors declare no competing interests.